\documentclass[twocolumn,showpacs,preprintnumbers,amsmath,amssymb]{revtex4}

\usepackage{graphicx}
\usepackage{dcolumn}
\usepackage{bm}

\begin{document}

\preprint{PRL Preprint}

\title{Controllable matter-wave switchers with vector Bose-Einstein solitons}
\author{Judit Babarro, Mar\'{\i}a J. Paz-Alonso, Humberto Michinel}
\affiliation{\'Area de \'Optica, Facultade de Ciencias de Ourense,\\
Universidade de Vigo, As Lagoas s/n, Ourense, ES-32004 Spain.}
 
\author{David N. Olivieri}
\affiliation{Escola T\'ecnica Superior de Enxeñer\'{\i}a Inform\'atica.\\ 
Universidade de Vigo, As Lagoas s/n, Ourense, ES-32004 Spain.}

\author{Jos\'e R. Salgueiro}
\affiliation{\'Area de \'Optica, Facultade de Ciencias de Ourense,\\
Universidade de Vigo, As Lagoas s/n, Ourense, ES-32004 Spain.}

\begin{abstract}
We show the possibility of producing matter-wave switching devices 
by using Manakov interactions between matter wave solitons in two-species 
Bose-Einstein Condensates (BEC). 
Our results establish the experimental parameters for three interaction 
regimes in two-species BECs: symmetric and asymmetric splitting, 
down-switching and up-switching. We have studied the dependence upon the 
initial conditions and the kind of interaction between the two components 
of the BECs.
\end{abstract}

\pacs{05.45.Yv, 42.65.Tg, 03.75.-b}

\maketitle
{\em Introduction.-} 
There has been remarkable experimental and theoretical progress in recent 
years regarding the interaction phenomena of coherent matter waves within 
Bose-Einstein condensates (BEC) of ultra cold atomic gases\cite{anderson95}. 
The interaction between the constitutive bosons inside the condensate is 
defined in terms of the ground state scattering length $a$. 
When $a>0$ the interaction 
between the particles in the condensate is repulsive, whereas for $a<0$, 
the interaction is attractive. Experimental preparation of negative scattering 
length was made possible by using Feshback resonances to continuously detune 
the value of $a$ from positive to negative values\cite{inouye98}. 
This provides new interest for analyzing systems of interacting condensates 
with an attractive coupling force, provided the number of particles is 
limited to avoid collapse \cite{gerton00-donley01}.  

In spite of this serious difficulty, negative scattering length condensates 
have some peculiarities which make them interesting. For instance, if 
the trap is removed in one direction and shrunk in the transverse plane, 
attractive interaction gives rise (for a given number of particles) to 
a self-confined stationary state\cite{perez-garcia98}. In this case, the cloud 
forms a soliton and can be controlled as a particle by acting on it with 
external fields\cite{khay02,strecker02}. 

On the other hand, results with two-species BECs\cite{myatt97} allow to 
perform experiments with two-component (vector) matter wave solitons. 
In this configuration, the interaction between the two species plays a 
significant role in determining the dynamics of the clouds\cite{modugno}. 
In this work, we extend the 
concept of Bose-Einstein soliton\cite{perez-garcia98,khay02,strecker02}  
to the case of two-species condensates. We show the possibility of producing 
{\em Manakov solitons}\cite{manakov74,soljacic03} with matter waves and to 
use them in the design of matter-wave switching devices. To this aim, 
we have studied numerically the interaction dynamics of two-species 
condensates within the confining cylindrical 
trap for different scattering length regimes and initial conditions.
 
The theoretical foundations of our work are presented in the first part of 
this paper, where we describe the model and the experimental 
configuration which has been assumed. We used a coupled set of  nonlinear 
Schr\"odinger equations, properly describing Manakov 
solitons\cite{malomed91-busch97-radhak97}, for modeling the time evolution of 
the initial wave functions before and after the interaction. 
Next, we demonstrate that some physical insight into the dynamics can be 
obtained from a set of equations derived directly from a variational 
analysis\cite{anderson88-perez-garcia97}, at least for the case of 
stationary states, while more complicated \textit{scattering} phenomena, 
must be calculated from direct numerical integration of the field equations. 
We show that interesting switching devices can be implemented with these 
interactions. Finally, we present our conclusions.

{\em Theoretical Foundations.-}
The system we have studied consists of two BEC matter waves of different 
species which interact via 2-body elastic effects. 
We have assumed an asymmetric {\em cigar} trapping potential in which, 
once individual BEC solitons are formed, they can be given an initial momentum 
along $x$ with an external force to make them collide. Assuming that the 
dynamics of the condensate is frozen in the transversal plane $y,z$ and the 
trap is switched off along the $x$ direction, the behavior of each condensate 
can be described through a one-dimensional Gross-Pitaevskii equation (GPE), 
with the addition of a coupling term that takes into account the interaction 
potential between the two condensate wave functions:

\begin{equation}
\label{eq:GPnorm}
i\frac{\partial \psi_k}{\partial\tau} + \frac{1}{2}\frac{\partial^2 \psi_k}{\partial\eta^2} 
+ ( a_{kk}|\psi_k|^2 + a_{jk} |\psi_{j}|^2 ) \psi_k=0,
\end{equation}
where $k=1,2$, $j=2,1$. We have defined the dimensionless variables 
$\tau=(\hbar/mL^2)t$ and $\eta=x/L$, together with the normalized wave function
(order parameter) $\psi_i(\eta,\tau)$, which gives the number of particles 
per unit length. Thus, the normalization for $\psi_k$ is 
$N_i = \int |\psi_i|^2 d\eta$, 
being $L\approx1\mu m$ the radial dimension of the trap, $a$ the ground 
state scattering length (typically $a\approx3a_0$, with $a_0$ the Bohr radius) 
and $m$ the mass of the atoms.

The analytic solution to the set of equations (\ref{eq:GPnorm}) represents 
a difficult problem. For some particular cases, explicit formulas were given 
by Manakov\cite{manakov74}. However, we can obtain approximate results by 
assuming an initial Gaussian wave function and performing a variational 
analysis\cite{anderson88-perez-garcia97}. This provides physical insight 
into the propagation and elastic collisions of fundamental parameters. 
Although this procedure breaks down for inelastic scattering when the 
beams split off, we will use these analytic results as a guide when 
we perform a full numerical integration of Eqs.(\ref{eq:GPnorm}).
To obtain the differential equations for the motion of the centroid 
and width of the initial Gaussian trial wave functions, we must minimize the 
Lagrange density, which produces the GP equation,
over a set of Gaussian trial wave functions such that: 

\begin{equation}
\label{uk}
\psi_k(\eta,\tau) = A_k \cdot \exp\biggl[ -\frac{(\eta-\eta_k)^2}{2\sigma_k^2} 
+ i\biggl(\eta\alpha_k + \eta^2\beta_k \biggr)  \biggr], 
\end{equation}
where $k=1,2$. Inserting this trial function into the Lagrangian density and 
taking the variation with respect to $\eta$, we obtain a set of 
differential equations with the above parameters, all of them $\tau$ 
dependent:  $A_{k}$ (complex amplitude), $\sigma_{k}$ 
(half width of the cloud), $\eta_{k}$ (position of the centroid), 
$\alpha_k$ (velocity) and $\beta_k$ (inverse square root of the beam 
curvature radius). 
The equations obtained describe the motion of the centroid and the 
oscillations of the soliton widths. Thus, taking the variation with respect 
to $A_{k}$ and $A^{*}_{k}$, and operating with them, we obtain the 
conservation of the number of particles: \(\dot N_k=0 \). 
Taking the variation with respect to the parameter $\alpha_k$ and 
defining $\eta_{jk}=\eta_j-\eta_k$ (distance between centroids), 
we obtain the evolution of the separation of the beam centers and widths:

\begin{equation}
\begin{split}
\label{ddrho}
\ddot\eta_{jk} = -\frac {2 a_{jk} (N_j+N_k) \eta_{jk}}{\pi(\sigma_j^2 + \sigma_k^2)^{3/2}} 
\exp\biggl[ - \frac{\eta_{jk}^2}{\sigma_j^2 + \sigma_k^2} \biggr],
\end{split}
\end{equation}

\begin{eqnarray}
\label{sigma1}
\ddot\sigma_j &=& \frac{1}{\sigma_j^3} -  \frac{a_{jj}N_j}{2\pi \sigma_j^2}\\ \nonumber 
&-& \frac{2 a_{jk} \sigma_j N_k}{\pi(\sigma_j^2 + \sigma_k^2)^{3/2}}
\biggl[ 1-\frac{2\eta_{jk}^2}{\sigma_j^2 + \sigma_k^2} \biggr] 
\exp\biggl[ - \frac{\eta_{jk}^2}{\sigma_j^2 + \sigma_k^2} \biggr].
\end{eqnarray}

Eq.(\ref{ddrho}) can be integrated to obtain the potential $\Pi_k$ 
ruling the interaction between the two species:

\begin{equation}
\label{potential}
\Pi_k = -\frac{a_{jk} N_j}{\pi \sqrt{\sigma_k^2 + \sigma_j^2}}
\exp\biggl[- \frac{\eta_{jk}^2}{\sigma_k^2 + \sigma_j^2} \biggr],
\end{equation}
with $k=1,2$; $j=2,1$. 
Equations (\ref{ddrho}) and (\ref{sigma1}), although not exact, are
valuable tools for a further detailed numerical exploration.
In first place, we must notice that depending on the sign 
of the coupling constant, the interaction potential will 
correspond to a barrier ($a_{jk}<0$) or to a well ($a_{jk}>0$). Thus, 
we can predict a minimum separation distance 
$\eta_{cr}$ as the half width estimate of $\Pi_k$, which corresponds 
to a value $\eta_{cr}\approx\sqrt{\sigma_j^2+\sigma_k^2}$. This means 
that two clouds separated more than their width will not interact. 
Our numerical simulations corroborate all these predictions from the 
variational analysis. 

On the other hand, when the two clouds are in the same 
initial position (\emph{i.e.}, $\eta_j(0) = \eta_k(0)$) with null velocities, 
the condition $\ddot\sigma_1 = \ddot\sigma_2 = 0$, predict the existence 
of stationary states corresponding to a zero spreading of the wave functions. 
These {\em vector solitons} are formed due to self-trapping from $a_{jj}$ 
and $a_{kk}$ terms in Eq. (\ref{eq:GPnorm}) together with cross-interaction 
($a_{jk}$). 
We have performed a numerical calculation by means of a relaxation method 
of these stationary states for different values of the scattering lengths. 
The results are in agreement with the variational calculations. 
As we will comment below, the computer simulations reveal that tunning 
the value and sign of the $a_{jk}$ can yield to a dramatic change in the 
stability of the stationary states. From a physical point of view, it is 
evident that a repulsive cross-interaction ($a_{jk}<0$) opposed to the 
self-trapping effect could yield to an instabilization of the vector BEC 
soliton. This situation is also predicted by the variational calculation 
when  $\ddot\sigma_j= 0$ in Eq. (\ref{sigma1}). In this case, the stationary 
states will only exist if:

\begin{equation}
\label{condition}
\biggl |\frac{a_{jk}}{a_{jj}}\biggr | \leqslant \frac{N_j}{N_k}\frac{(1+\sigma_k^2 /\sigma_j^2)^{3/2}}{2\sqrt 2},
\end{equation}
For the particular case of two condensates with equal number of particles 
and widths, the above condition (\ref{condition}) takes the form: 
$|a_{jk}/a_{jj}| \leqslant 1$. The numerical simulation sets this value to
$|a_{jk}/a_{jj}| \leqslant 0.95$, which is in very good agreement with the 
theoretical prediction.

{\em Numerical simulations.-}
For the integration of Eqs.(\ref{eq:GPnorm}), we utilized a 
Crank-Nicholson finite difference discretization with 500 points 
grid. Our simulations consisted, in first place, of comparing different 
velocities and number of particles of the incident BEC soliton wave 
functions described, and observing the evolution of the clouds after 
the interaction, for several scattering lengths. From an exhaustive 
numerical exploration, we realized that there exist well defined 
scattering regimes for which the emerging wave functions can be radically 
different. We also explored the effect of tunning the value of the scattering 
length for several collision processes. We have found that a change 
of the sign and the values of the $a_{jj}$ and $a_{jk}$ terms has a 
deep influence on the dynamics of the clouds. 

\begin{figure}
{\centering \resizebox*{1\columnwidth}{!}{\includegraphics{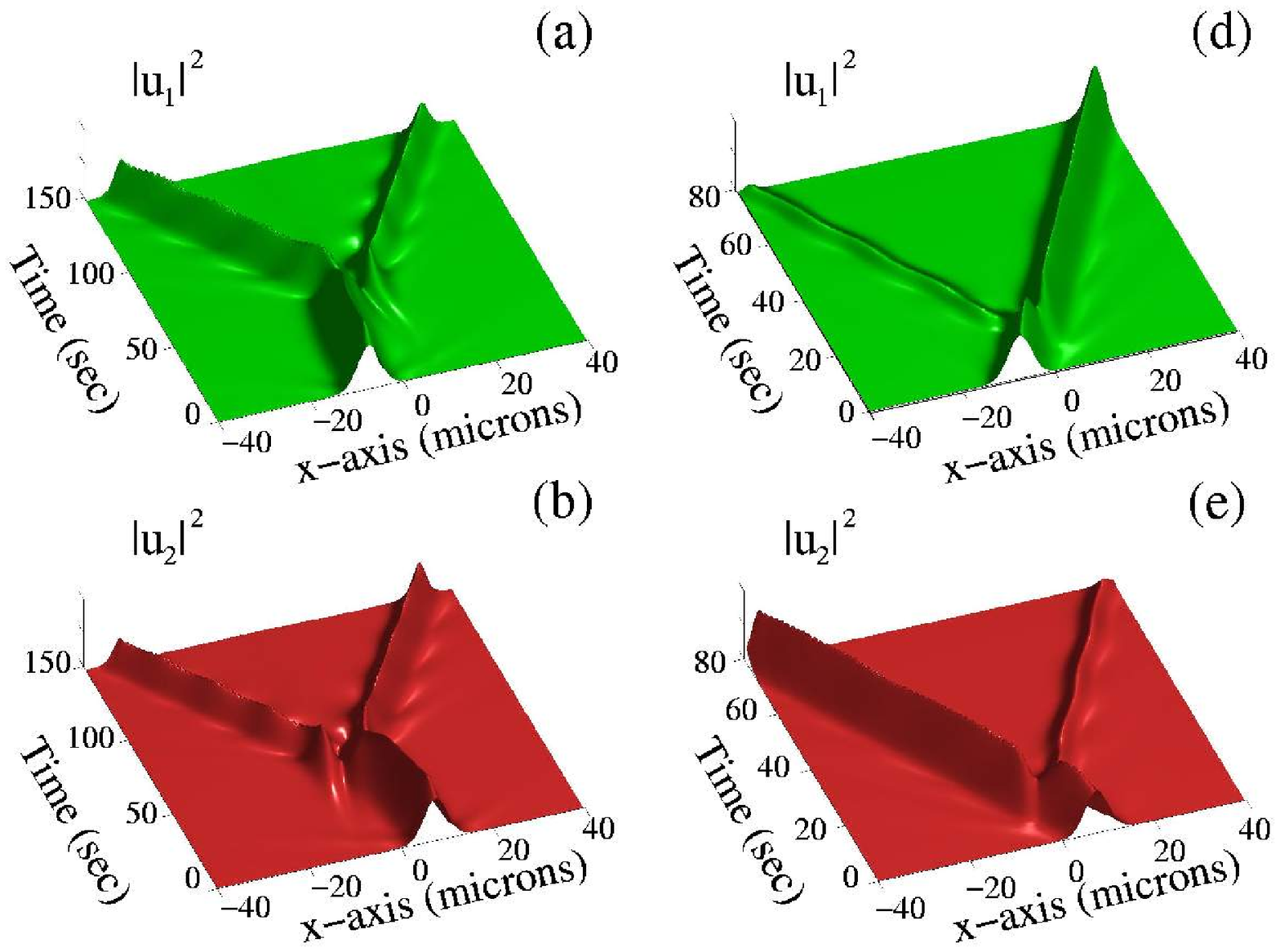}} \par}
{\centering \resizebox*{1\columnwidth}{!}{\includegraphics{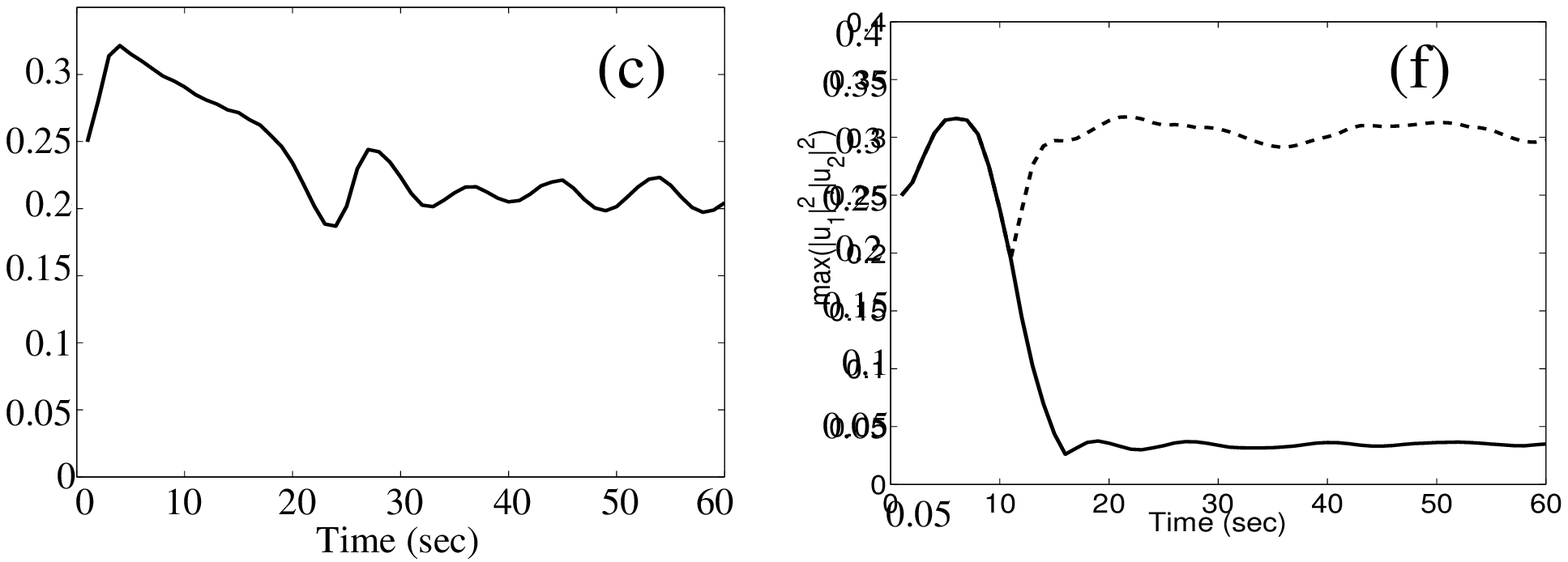}} \par}
\caption{
[Color on line] Different Y-switching configurations based upon the 
characteristics of the incoming wave functions. The simulations correspond
to the parameters: $N_1=N_2=4000$, width of the clouds $10\mu m$,
initial separation of the centroids $x_0=15\mu m$, and
initial velocities $v_{i1}=-v_{i2}=0.05\mu m/s$ (left), and
$v_{i1}=-v_{i2}=0.5\mu m/s$ (right). Scattering length parameters
are: $a_{11}=a_{22}=1.0$ and $a_{12}=a_{21}=2.0$.
\label{figzonas}}
\end{figure}

Thus, we will show in first place the results obtained for the case of 
attractive scattering lengths coefficients (i.e.: $a_{jj}$ and $a_{jk}>0$). 
Fig. \ref{figzonas} illustrates different Y-switching 
processes based upon the characteristics of the incoming wave function. 
In (a) and (b) we show respectively the square modulus of $u_1$ and $u_2$. 
The simulations correspond to the parameters: $N_1=N_2=4000$, width 
of the clouds $w_1=w_2=10\mu m$, initial separation of the centroids 
$x_0=15\mu m$, and initial velocities $v_{i1}=-v_{i2}=0.05\mu m/s$. 
The scattering length adimensional coefficients are $a_{11}=a_{22}=1.0$ 
and $a_{12}=a_{21}=2.0$. The two incoming clouds split symmetrically into 
two mutually trapped {\em vector} BEC solitons with the same number of 
particles. In c) the maximum of $|u_1|^2$ and $|u_2|^2$ are plotted. 
Figs. \ref{figzonas}-d and \ref{figzonas}-e
correspond to the same parameters but increasing the initial velocities one
order of magnitude: $v_{i1}=-v_{i2}=0.5\mu m/s$. In this case the emerging 
distributions are not symmetric, as can be clearly appreciated in 
Fig. \ref{figzonas}-f where the evolution of both peaks of 
$|u_1|^2$ is shown. 
In both cases the resulting matter waves are stable vector BEC solitons
with larger velocities than the input beams. The wave functions take the form
$\alpha_1u_1+\alpha_2u_2$, being $|\alpha_1|^2+|\alpha_2|^2=1$. We have 
found numerically that the different splitting regimes are limited by 
the following values: $0<|v_i|<0.1\mu m/s$ for the symmetric case and
$0.1\mu m/s<|v_i|<0.9\mu m/s$, for the splitting with different number of 
particles. For initial velocities $|v_i|>0.9\mu m/s$ the two wave functions 
are mutually transparent and arise with any change after the collision.

\begin{figure}
{\centering \resizebox*{1\columnwidth}{!}{\includegraphics{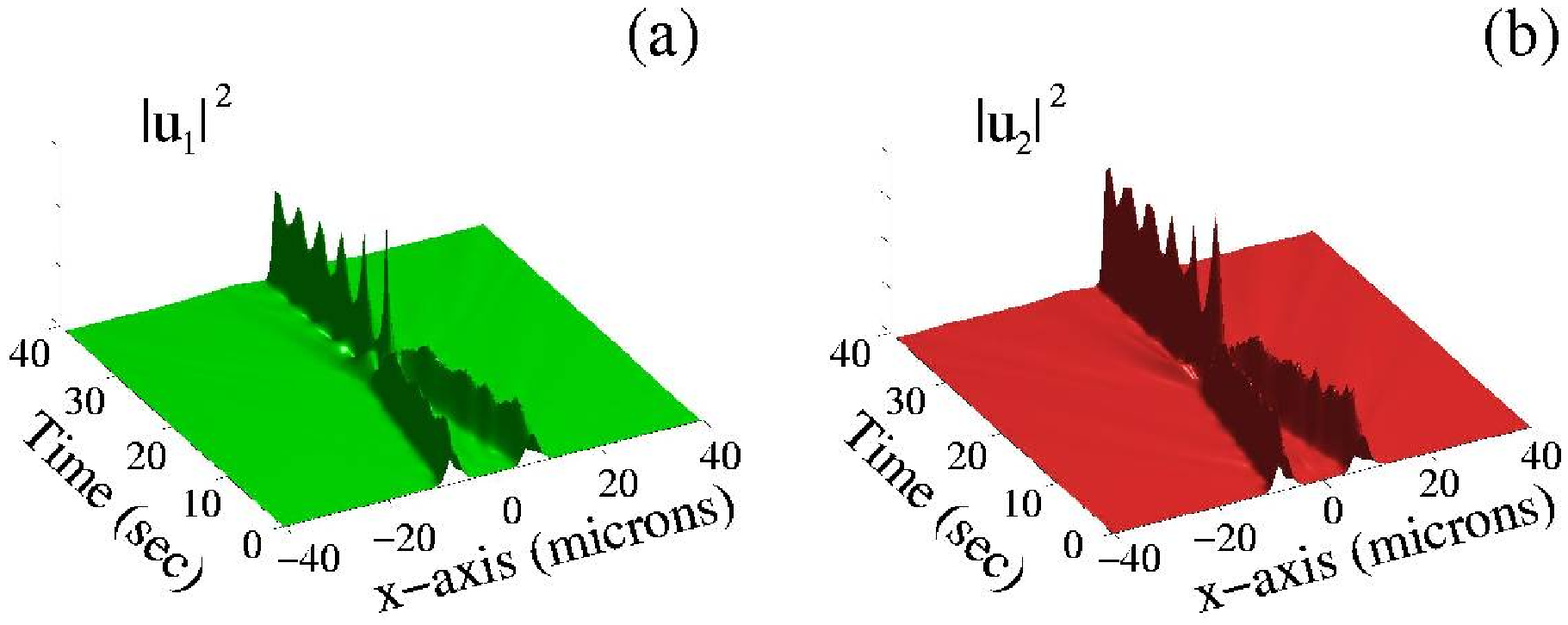}} \par}
{\centering \resizebox*{1\columnwidth}{!}{\includegraphics{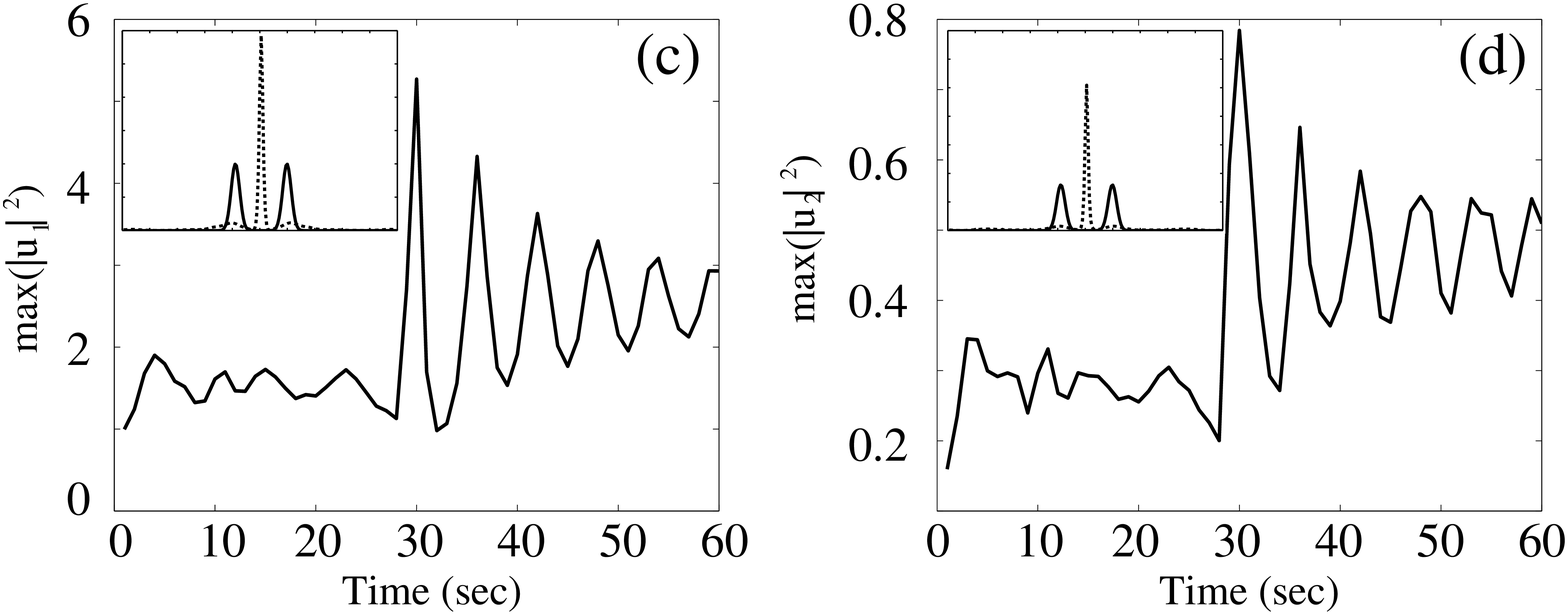}} \par}
\caption{
[Color on line] Up-switching of two condensate pairs with different number 
of particles.
$N_{1l}=N_{1r}=8000$ (a) and $N_{2l}=N_{2r}=3000$ (b);
widths $5\mu m$; initial velocities $v_{1}=-v_{2}=0.3\mu m/s$ and 
initial separation of the centroids $x_0=15\mu m$.  
Scattering length coefficients: $a_{11}=a_{22}=1.0$ and  $a_{12}=a_{21}=2.0$. 
(c) and (d) show the evolution of the peaks of (a) and (b), respectively. 
Insets show input and output wavefuntions for the respective cases.
\label{fusion}}
\end{figure}

The collision of two vector BEC solitons constructed by superposition
of Gaussian-shaped clouds is shown in Figure \ref{fusion}. 
In this case it is obtained an up-switching of the peak of both wave functions
which emerge fused after the collision. The simulation corresponds to 
the parameters: number of particles: $N_{1l}=N_{1r}=8000$ (a) and 
$N_{2l}=N_{2r}=3000$, where subindices $l,r$ designate the soliton coming form
left or right side (b); cloud widths $5\mu m$; initial velocities 
$v_{l}=-v_{2}=0.3\mu m/s$ and initial separation of the centroids 
$x_0=15\mu m$.
The scattering length coefficients are: $a_{11}=a_{22}=1.0$ and  
$a_{12}=a_{21}=2.0$. We must stress that the fusion of two condensates 
only takes place for clouds with different number of particles.

\begin{figure}
{\centering \resizebox*{1\columnwidth}{!}{\includegraphics{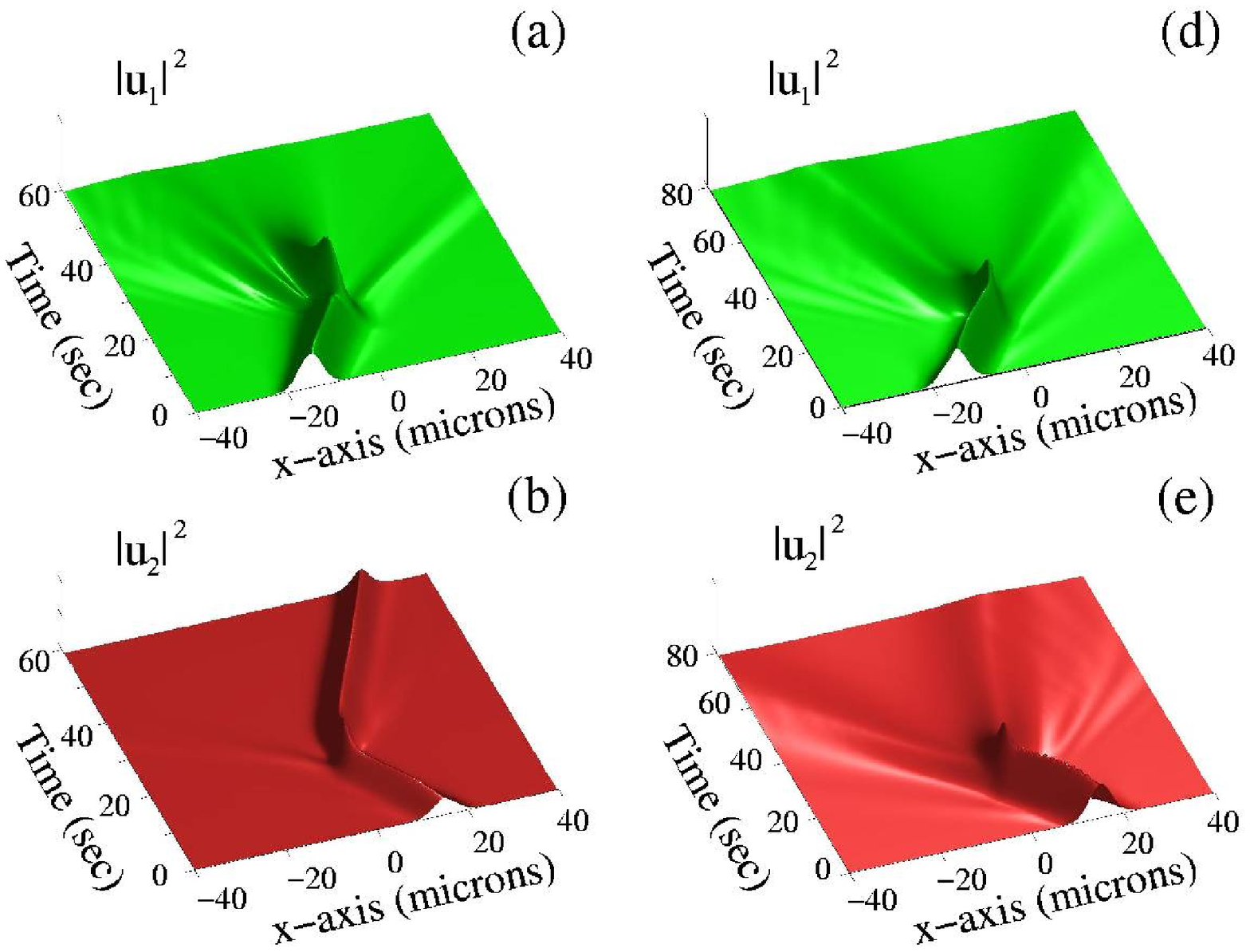}} \par}
{\centering \resizebox*{1\columnwidth}{!}{\includegraphics{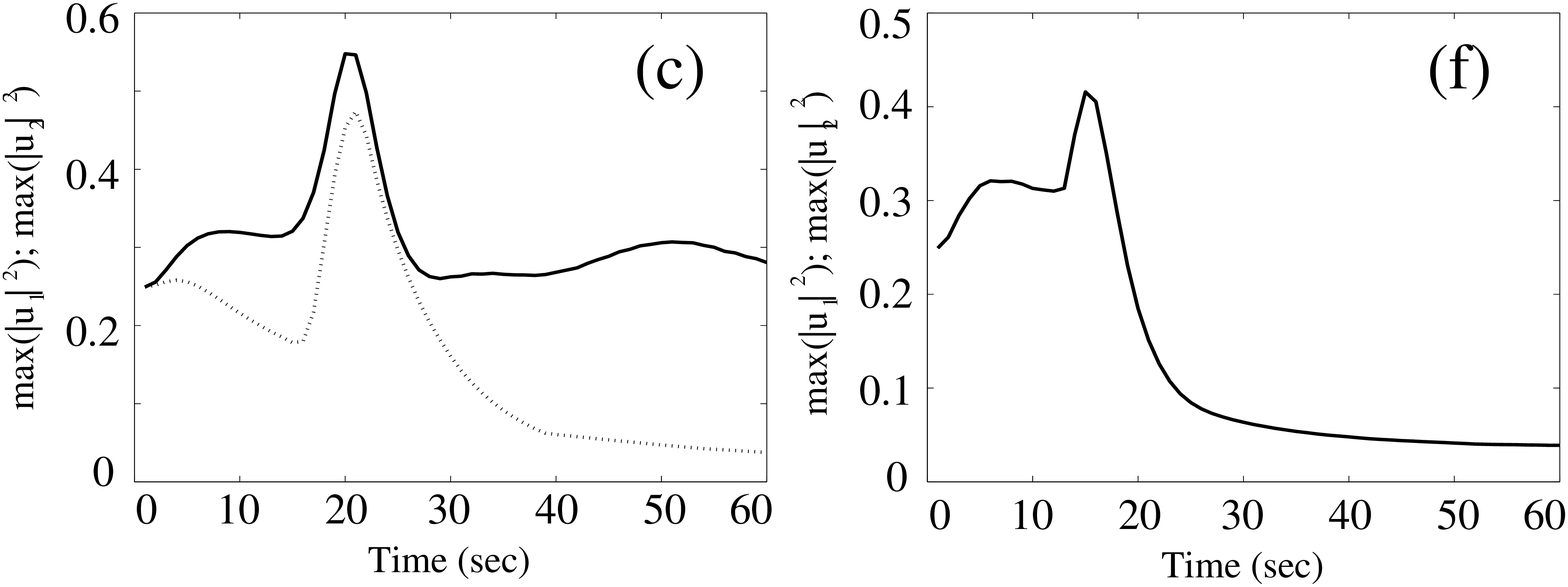}} \par}
\caption{
[Color on line] Down-switching of one condensate (left) and two condensates 
(right), in the case of repulsive cross-interaction. 
The simulations correspond to the parameters: 
number of particles $N_1=N_2=4000$, width of the clouds $10.0\mu m$,       
initial separation of the centroids $x_0=30.0\mu m$,
and initial velocities $v_{1}=v_{2}=0.8\mu m/s$.
The scattering length parameters are:
$a_{11}=0.6$, $a_{22}=1.0$, and $a_{12}=a_{21}=-2.3$ (left);
$a_{12}=a_{21}=-1$ and $a_{11}=a_{22}=1.0$ (right). 
Peak power plots of the condensates before and after 
the collision are shown in (c) and (f) for respective cases. 
\label{fignegcoup}}
\end{figure}

We will now consider the case of repulsive cross-interaction ($a_{jk}<0$). 
This situation is very interesting, as it corresponds to condensates of 
different types of atoms (for instance $^{85}Rb$ and $^{40}K$) as in the 
experiments from ref.\cite{modugno}. As it can be appreciated in  
Fig. \ref{fignegcoup}, the results of the collisions are  very different 
comparing with Fig. \ref{figzonas}, which corresponds to attractive
cross-interaction. Figs. \ref{fignegcoup} (a) and (b) show the down-switching 
of one of the condensates which is almost annihilated by the other, which is 
repelled. The plot (c) shows the evolution of the peaks of both distributions. 
The parameters used for the calculation are typical of experiments:
$N_1=N_2=4000$, widths of the clouds $5.0\mu m$, initial separation of the 
centroids $x_0=30\mu m$, and input velocities $v_{1}=-v_{2}=0.8\mu m/s$. The 
values of adimensional scattering length parameters $a_{11}=0.6$, $a_{22}=1.0$, 
and $a_{12}=a_{21}=-2.3$ were taken form ref. \cite{modugno}.
In figures \ref{fignegcoup}-d and \ref{fignegcoup}-f we show the same 
simulation but taking $a_{11}=a_{22}=1.0$, and $a_{12}=a_{21}=-1$ 
In this case the effect of the collision is dramatic and both clouds almost 
completely spread.

\begin{figure}
{\centering \resizebox*{1\columnwidth}{!}{\includegraphics{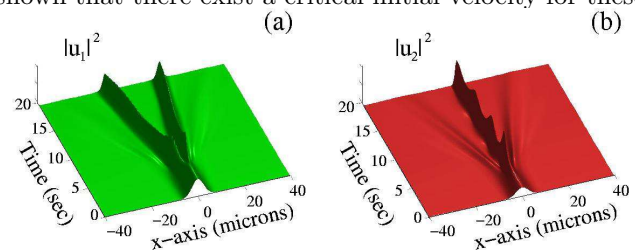}} \par}
\caption{
[Color on line] Propagation of two initially superposed 
condensates in the case of repulsive cross-interaction,  with 
a slightly different number of particles $N_{1}=15430$ and $N_{2}=15750$. 
The widths of the clouds are $10\mu m$ and the scattering length parameters 
are $a_{11}=a_{22}=1.0$ and $a_{12}=a_{21}=-0.8$.
\label{div}}
\end{figure}

Finally we have studied the evolution of two initially superposed
condensates in the case of repulsive cross-interaction, 
with $a_{11}=a_{22}=1.0$ and $a_{12}=a_{21}=-0.8$. The widths of the clouds 
are $10\mu m$ and $N_{1}=N_{2}=15500$ have been chosen to satisfy the 
condition of stationary state from Eq. (\ref{condition}). We have observed 
that for clouds with almost exact number of particles, the existence of the 
stationary state is not affected by the negative sign of $a{jk}$. However, 
if the number of particles is slightly altered (taking, for instance 
$N_{1}=15430$ and $N_{2}=15750$), the condensate with less population splits 
off in two solitons and the other condensate remains oscillating, as shown 
in Fig.\ref{div}.

{\em Conclusions.-}
We have studied, through an approximate theoretical derivation and through 
numerical simulations, several consequences of the interaction of composite 
Bose Einstein mutually interacting solitons. These systems give rise to a 
mutual trapping phenomena due to the soliton-soliton interaction forces. 
As a result of these interactions, we have predicted new phenomena not 
previously detected for the formation of stable vector matter-wave solitons, 
after a violent division, and have shown that there exist a critical 
initial velocity for these phenomena to occur. Depending on the values 
and signs of the scattering length coefficients,
we have found several different switching regimes for matter wave solitons. 
Finally, we have calculated the effect of a repulsive cross-interaction in 
the stability of vector Bose-Einstein solitons.


\begin{thebibliography}{10}
\bibitem{anderson95} M. H. Anderson, J. R. Ensher, M. R. Matthews, C. E. 
Wieman, and E. A. Cornell, Science \textbf{269}, 198-201 (1995).
\bibitem{inouye98} S. Inouye, M.R. Andrews, J. Stenger, H.-J. Miesner, 
D.M. Stamper-Kurn, and W. Ketterle, Nature \textbf{392}, 151-154 (1998). 
\bibitem{gerton00-donley01} J. M. Gerton, D. Strekalov, I. Prodan, and 
R. G. Hulet, Nature \textbf{408}, 692-695, (2000);
E.A. Donley, N.R. Claussen, S.L. Cornish, J.L. Roberts, 
E.A. Cornell and C.E. Wieman, Nature \textbf{412}, 295-299, 2001.
\bibitem{perez-garcia98} V. M. P\'erez-Garc\'ia, H. Michinel, 
and H. Herrero, Phys. Rev. A \textbf{57}, 3837 (1998).
\bibitem{khay02} L. Khaykovich, F. Schreck, G. Ferrari, T. Bourdel,
J. Cubizolles, L. D. Carr, Y. Castin, and C. Salomon, Science \textbf{296},
1290-1293, (2002).
\bibitem{strecker02} K. E. Strecker, G. B. Partridge, A. G. Truscott, and
R. G. Hulet, Nature \textbf{417}, 150-153, (2002).
\bibitem{myatt97} C.J. Myatt, E.A. Burt, R.W.Ghrist, E.A. Cornell, C.E. 
Wieman, Phys. Rev. Lett. \textbf{78}, 586-589, (1997).
\bibitem{modugno} G. Modugno, M. Modugno, F. Riboli, G. Roati, and
M. Inguscio, Phys. Rev. Lett. \textbf{89}, 190404 (2002);
F. Riboli and M. Modugno, Phys. Rev. A \textbf{65}, 063614 (2002).
\bibitem{manakov74} S. V. Manakov, Sov. Phys.-JETP \textbf{38}, 248-252, 
(1974).
\bibitem{soljacic03} M. Soljacic, K. Steiglitz, S. M. Sears, M. Segev,
M. H. Jakubowski, and R. Squier, Phys. Rev. Lett. \textbf{90},
254102-1-4 (2003).
\bibitem{malomed91-busch97-radhak97} B. A. Malomed and S. Wabnitz, 
Opt. Lett. \textbf{16}, 1388-1390, (1991);
T. Busch, J. I. Cirac, V. M. Perez-Garcia, and P. Zoller,
Phys. Rev. A \textbf{56}, 2978-2983 (1997);
R. Radhakrishnan, M. Lakshmanan, and J. Hietarinta, 
Phys. Rev. E \textbf{56}, 2213-2216, (1997).
\bibitem{anderson88-perez-garcia97} D. Anderson, M. Lisak, and T. Reichel, 
Phys. Rev. A \textbf{38}, 1618-1620, (1988);
V. M. P\'erez-Garc\'ia, H. Michinel, J. I. Cirac, 
M. Lewenstein, and P. Zoller, 
Phys. Rev. A \textbf{56}, 1424-1434, (1997).
\end{thebibliography}
\end{document}